\documentstyle[12pt]{article}
\renewcommand{\theequation}{{\rm
\thesection.\arabic{equation}}}
\newcommand{\be}{\begin{equation}}
\newcommand{\ee}{\end{equation}}
\newcommand{\bea}{\begin{eqnarray}}
\newcommand{\eea}{\end{eqnarray}}
\newcommand{\nn}{\nonumber}

\newcommand{\ketone}{|1\rangle}
\newcommand{\kettwo}{|2\rangle}

\newcommand{\hx}{\hat{r}}

\newcommand{\ha}{\hat{a}}
\newcommand{\hq}{\hat{q}}
\newcommand{\had}{\hat{a}^\dagger}
\newcommand{\hH}{\hat{H}}
\newcommand{\redavone}{\langle 1 \|r{\bf C}^{(1)}\| 2 \rangle}
\newcommand{\redavtwo}{\langle 1\|r^2{\bf C}^{(2)}\| 2 \rangle}
\newcommand{\Abarone}{\bar{A}^{(E1)}_{12}}
\newcommand{\Abartwo}{\bar{A}^{(E2)}_{12}}
\newcommand{\Aone}{A^{(E1)}_{12}}
\newcommand{\Atwo}{A^{(E2)}_{12}}

\newcommand{\br}{{\bf x}}
\newcommand{\bc}{{\bf c}}
\newcommand{\bn}{{\bf n}}

\newcommand{\beps}{{\bf \epsilon}}
\newcommand{\bb}{{\bf b}}
\newcommand{\ket}[1]{{|#1\rangle}}
\newcommand{\bra}[1]{{\langle#1|}}

\newcommand{\average}[3]{{\langle#1|#2|#3\rangle}}

\newcommand{\threej}[6]{\left(\begin{array}
{ccc}#1&#2&#3\\#4&#5&#6\end{array}\right)}

\begin{document}
\baselineskip 3.4ex

\begin{center}
{\Huge \bf Quantum dynamics of cold trapped ions, with
application to quantum computation}

\bigskip\bigskip
{\small by }
 
{Daniel F. V. James}\\

\bigskip\bigskip

{\small Theoretical Division (T--4)}\\
{\small Mailstop B-268}\\
{\small Los Alamos National Laboratory}\\
{\small Los Alamos, NM 87545}\\
{\small U.S.A.}\\
{\small TEL: (505)-667-0956}\\
{\small FAX: (505)-665-3909}\\
{\small email: dfvj@t4.lanl.gov}\\

\bigskip
{to be submitted to \it Applied Physics B: Lasers and Optics}\\
\bigskip\bigskip
{\bf Abstract}\\
\end{center}
The theory of interactions between lasers and 
cold trapped ions as it pertains to the design of
Cirac-Zoller quantum computers is discussed.  
The mean positions of the trapped ions, the
eigenvalues and eigenmodes of the ions' oscillations,
the magnitude of the Rabi frequencies for both
allowed and forbidden internal transitions of the ions and
the validity criterion for the required Hamiltonian are
calculated.   Energy level data for a variety of
ion species is also presented.
\bigskip
\begin{center}

 \end{center}
\newpage

\section{Introduction}
\setcounter{equation}{0}

 A quantum computer is a
device in which data can be stored in a network of quantum mechanical
two-level systems, such as spin-1/2 particles or two level atoms.  The
quantum mechanical nature of such systems allows the possibility of a 
powerful new feature to be incorporated into data processing, namely
the capability of performing logical operations upon quantum mechanical
 superpositions of numbers. Thus in a conventional digital computer 
each data register is, throughout any computation, always in a definite 
state ``1'' or ``0''; however in a quantum computer, if such a device can be
realized, each data register (or ``qubit'') will be in an undetermined 
quantum superposition of two states $\ket{1}$ and
$\ket{0}$.  Calculations would then be performed by external interactions
with the various two-level systems that constitute the device, in such a way
 that conditional gate operations involving two or
more different qubits can be realized. The final result would be 
obtained by measurement of the quantum mechanical
probability amplitudes at the conclusion of the calculation. 
Much of the recent interest in practical quantum computing
has been stimulated by the discovery of  a quantum algorithm
which allows the determination of the prime factors
of large composite numbers efficiently \cite{Shor},
and of coding schemes that, provided 
operations on the qubits can be performed within a certain 
threshold degree of accuracy, will allow arbitarily complicated
quantum computations to be performed reliably {\it regardless
of operational error} \cite{MannyRay}. 

So far, the most promising hardware proposed for implementation
of such a device seems to be the cold-trapped ion system devised 
by Cirac and Zoller \cite{CZ}.  Their design, which is
shown schematically in figure 1, consists of a string
of ions stored in a linear radio-frequency trap and cooled sufficiently
that their motion, which is coupled together due to the Coulomb force
between them, is quantum mechanical in nature. Each qubit would be
formed by two internal levels of each ion, a laser being used to perform
manipulations of the quantum mechanical probability amplitudes of the
states; conditional two-qubit logic
gates being realized with aid of the excitation or de-excitation of quanta of
the ions' collective motion.  For a more detailed description of the
concept of cold-trapped ion quantum computation, the reader is referred
to the article by Steane \cite{Steane}.

There are two distinct possibilities for the
choice of the internal levels of the ion: firstly the two states could be
the ground state, and a metastable excited state of the ion (or
more precisely, sublevels of these states); secondly
the two states could be two nearly degenerate sublevels of the ground 
state.  In the first case a single laser would suffice to perform the required
operations; in the second, two lasers would be required to perform Raman
transitions between the states, via a third level.  Both of these schemes
have advantages: the first, which I will refer to as the
``single photon'' scheme, has the great advantage of conceptual 
and experimental simplicity; the second,
the ``Raman scheme'',
offers the advantages of a very low rate for spontaneous
decay between the two nearly degenerate states and resilience 
against fluctuations of the phase of the laser. This later scheme
was recently used by the group headed by Dr. D. J. Wineland
at the National Institute of Science and Technology at Boulder,
Colorado to realize a quantum logic gate using a {\it single}
trapped Beryllium ion \cite{NIST}.

In this paper I will discuss the theory of laser interactions with
cold trapped ions as it pertains to the design of a Cirac-Zoller
quantum computer.  I will conscentrate on the ``single photon
scheme'' as originally proposed by those authors, although
much of the analysis is also relevant to the ``Raman  scheme''.
Fuller accounts of aspects of this are availible 
in the literature: see for example \cite{WI}, \cite{CBZP}, \cite{Steane};
however the derivation of several results are presented here for the
first time.   I will also present relevant data
gleaned from various sources
on some species of ion suitable for use in a quantum computation.

\section{Equilibrium positions of ions in a linear trap}\setcounter{equation}{0}

Let us consider a chain of $N$ ions in a trap.  The
ions are assumed to be strongly bound in the $y$ and $z$
directions but weakly bound in an harmonic potential in the 
$x$ direction. The
position of the $m-th$ ion, where the ions are numbered from left 
to right, will be denoted $x_m(t)$.  The motion of each ion will 
be influenced by
an overall harmonic potential due to the trap electrodes,
and by the Coulomb force exerted
by all of the other ions.  Hence the potential energy of 
the ion chain is given by the following expression:
\be
V=\sum^N_{m=1}\frac{1}{2}M\nu^2x_m(t)^2+
\sum_{\stackrel{\scriptstyle n,m=1}{m \neq n}}^N
\frac{Z^2e^2}{8\pi\epsilon_0}\frac{1}{|x_n(t)-x_m(t)|} \ ,
\ee
where $M$ is the mass of each ion,  $e$ is the electron charge, $Z$ is the
degree of ionization of the ions, $\epsilon_0$ is the permitivity of free space 
and $\nu$ is the trap frequency, which characterizes the strength of the trapping
potential in the axial direction.  Note that this is an 
unconventional use of the symbol $\nu$, which
often denotes frequency rather than angular frequency; 
following Cirac and Zoller, I will use
$\omega$ to denote the angular frequencies of the laser or the
transitions between internal states of the ions, and $\nu$
to denote angular frequencies associated with the motion of the
ions.

Assume that the ions are sufficiently cold
that the position of the $m-th$ ion can be approximated by
the formula
\be
x_m(t)\approx x_m^{(0)}+q_m(t)
\ee
where $x_m^{(0)}$ is the equilibrium position of
the ion, and $q_m(t)$ is a small displacement. The
equilibrium positions will be determined by the following
equation:
\be
\left[\frac{\partial V}{\partial x_m}\right]_{x_m=x_m^{(0)}}=0 
\label{equileqn1}
\ee
If we define the length scale $\ell$ by the formula
\be
\ell^3=\frac{Z^2e^2}{4\pi\epsilon_0 M\nu^2} \ ,
\label{eleanor}
\ee
and the dimensionless equilibrium position 
$u_m=x_m^{(0)}/\ell$, then
(\ref{equileqn1}) may be rewritten as
the following set of $N$ coupled algebraic equations
for the values of $u_m$:
\be
u_m-\sum_{n=1}^{m-1}\frac{1}{(u_m-u_n)^2}+
\sum_{n=m+1}^{N}\frac{1}{(u_m-u_n)^2}=0
 \quad (m=1,2,..N) \ .
\label{ionequs}
\ee
For $N=2$ and $N=3$ these equations may be solved
analytically:
\bea
N=2:&\quad&u_1=-(1/2)^{2/3}, 
\quad u_2=(1/2)^{2/3}\ , \\
N=3:&\quad&u_1=-(5/4)^{1/3}, 
\quad u_2=0 , 
\quad u_3=(5/4)^{1/3}\ . 
\eea
For larger values of $N$ it is necessary to solve for the values
of $u_m$ numerically.  The numerical  values of the solutions
to these equations for  2 to 10 ions is given in table 1.  Determining 
the solutions for larger numbers of ions is a straightforward but
time consuming task.

By inspection, the minimum value
of the spacing between two adjacent ions occurs at the center of the
ion chain.  Compiling the numerical data for the minimum
value of the separation for different numbers of trapped ions,
we find that it obeys the following relation:
\be
u_{min}(N) \approx \frac{2.018}{N^{0.559}} \ . \\
\label{separations}
\ee
This relation is illustrated in figure 2.  Thus the minimum
inter-ion spacing for different numbers of ions is given by the
following formula:
\be
x_{min}(N)=\left(\frac{Z^2e^2}
{4\pi\epsilon_0 M\nu^2}\right)^{1/3}
\frac{2.018}{N^{0.559}} \ . \\
\ee
This relationship is  important in determining the
capabilities of cold trapped ion quantum computers \cite{HJKLP}.

\section{Quantum Fluctuations of the ions}\setcounter{equation}{0}

This section discusses the equations of motion 
which describe
the displacements of the ions from their equilibrium positions.
Because of the Coulomb interactions between the ions,
the displacements of different ions will be 
coupled together. The Lagrangian
describing the motion is then
\be
L=\frac{M}{2}\sum^N_{m=1}(\dot{q}_m)^2
-\frac{1}{2}\sum^N_{n,m=1}q_nq_m
\left[\frac{\partial^2V}{\partial x_n \partial x_m}\right]_0
\ee
where the subscript $0$ denotes that the double partial derivative
is evaluated at $q_n=q_m=0$.  The partial derivatives may be
calculated explicitly to give the following expression:
\be
L=\frac{M}{2}\left[\sum^N_{m=1}(\dot{q}_m)^2
-\nu^2\sum^N_{n,m=1}A_{nm}q_nq_m
\right]
\label{lag1}
\ee
where 
\be
{\everymath {\displaystyle}
A_{nm}=\left\{\begin{array}{ll}
		1+2\sum^N_{\stackrel{\scriptstyle p=1}{p \neq m}}
		\frac{1}{\left|u_m-u_p\right|^3} & {\rm if}\quad n=m, \\
		\rule{0in}{5ex}
		\frac{-2}{\left|u_m-u_n\right|^3} & {\rm if}\quad n\neq m.
		\end{array}
	\right.}
\label{alice}
\ee
Since the matrix $A_{nm}$ is  real, 
symmetric and non-negative definite, its
eigenvalues must be non-negative.  The eigenvectors 
$b^{(p)}_m$ $(p=1,2,\cdots N)$
are therefore defined by the following formula:
\be
\sum^N_{n=1}A_{nm}b^{(p)}_n=\mu_pb^{(p)}_m 
\mbox{  } (p=1,\ldots, N) \ ,
\ee
where  $\mu_p\geq 0 $.  The eigenvectors  are assumed to
be numbered in order of increasing eigenvalue and to be 
properly normalized so that
\bea
\sum^N_{p=1}b^{(p)}_nb^{(p)}_m&=&\delta_{nm} \\
\sum^N_{n=1}b^{(p)}_nb^{(q)}_n&=&\delta_{pq} \ .
\label{beeorth}
\eea

The first eigenvector (i.e. the eigenvector with the smallest eigenvalue)
can be shown to be
\be
\bb^{(1)}=\frac{1}{\sqrt{N}}\left\{1,1,\ldots,1\right\}
{\rm ,}\qquad \mu_1=1 \ .
\label{beeone}
\ee
The next eigenvector can be shown to be
\be
\bb^{(2)}=\frac{1}{\left({\displaystyle \sum^N_{m=1}u_m^2}\right)^{1/2}}
\left\{u_1,u_2,\ldots,u_N\right\}{\rm ,}\qquad \mu_2=3 \ .
\ee
Higher eigenvectors must, in general, be determined numerically.
Equations (\ref{beeorth}) and (\ref{beeone}) imply that
\be
\sum^N_{m=1}b^{(p)}_m=0\quad{\mbox  {\rm if}}\: p\neq 1 .
\ee

For $N=2$ and $N=3$, the eigenvectors and eigenvalues
may be determined algebraically:
\bea
N=2:\quad \bb^{(1)}&=&\frac{1}{\sqrt{2}}(1,1),\quad\mu_1=1\ ,\nn \\
\bb^{(2)}&=&\frac{1}{\sqrt{2}}(-1,1),\quad \mu_2=3\ ,\nn \\
&&\\
N=3:\quad \bb^{(1)}&=&\frac{1}{\sqrt{3}}(1,1,1),\quad\mu_1=1\ ,\nn \\
\bb^{(2)}&=&\frac{1}{\sqrt{2}}(-1,0,1),\quad \mu_2=3\ ,\nn \\
\bb^{(3)}&=&\frac{1}{\sqrt{6}}(1,-2,1),\quad \mu_3=29/5 \ , 
\eea
For larger values of $N$, the eigenvalues and eigenvectors must
be determined numerically; their numerical values for  2 to 10 ions 
are given in table 2.

The {\em normal modes} of the ion motion are defined 
by the formula
\be
Q_p(t)=\sum^l_{m=1}b^{(p)}_mq_m(t) \ .
\ee
The first mode $Q_1(t)$ corresponds to all of the
ions oscillating back and forth as if they were rigidly
clamped together;  this is referred to as the {\em center of mass}
mode.  The second mode $Q_2(t)$ corresponds to each
ion oscillating with an amplitude proportional to its
equilibrium distance form the trap center;  This is called
the {\em breathing mode}.  
The Lagrangian for the ion oscillations (\ref{lag1})
may be rewritten in terms of these normal modes as follows:
\be
L=\frac{M}{2}\sum^N_{p=1}\left[\dot{Q}^2_p
-\nu_p^2Q^2_p\right] \ ,
\ee
where the angular frequency of the
$p-th$ mode is defined by
\be
\nu_p=\sqrt{\mu_p}\nu .
\label{nup}
\ee

This expression implies that the modes $Q_p$ are
uncoupled.  Thus the canonical momentum conjugate to
$Q_p$ is $P_p=M\dot{Q}_p$ and one can immediately
write the Hamiltonian as
\be
\hH=\frac{1}{2M}\sum^N_{p=1}P^2_p+
\frac{M}{2}\sum^N_{p=1}\nu_p^2 Q^2_p \ .
\ee
The quantum motion of the ions can now be considered
by introducing the operators \footnote{There is some arbitarinees
in the definition of the operators $\hat{P}_p$ and $\hat{Q}_p$,
which is related to the arbitrariness of the phase of the Fock states.  
I have used the definitions given by Kittel (ref. \cite{QTS}, p.16),
which differs from that given in other texts on quantum mechanics
(see for example, ref. \cite{Schiff}, p. 183 or ref.\cite{Milonni}
p. 36).}
\bea
Q_p\rightarrow\hat{Q}_p&=&i\sqrt{\frac{\quad\hbar\quad}
{2M\nu_p}}
\left(\ha_p-\had_p\right) \ , \\
P_p\rightarrow\hat{P}_p&=&\sqrt{\frac{\hbar M\nu_p}{2}}
\left(\ha_p+\had_p\right) \ .
\eea
where $\hat{Q}_p$ and $\hat{P}_p$ obey the canonical commutation
relation $\left[\hat{Q}_p,\hat{P}_q\right]=i\hbar\delta_{pq} $ and 
the creation and 
annihilation operators $\had_p$ and $\ha_p$ obey the
usual commutation relation $\left[\ha_p,\had_q\right]=\delta_{pq} $. 
  
Using this notation, the  interaction picture operator for the 
displacement of the 
$m-th$ ion from its equilibrium position is given by the formula:
\bea
{\hat q}_m(t)&=&\sum_{p=1}^Nb^{(p)}_m \hat{Q}_p(t) \nn \\
&=& i\sqrt{\frac{\hbar}{2M\nu N}}
\sum^N_{p=1}s^{(p)}_m
\left(\ha_pe^{-i\nu_pt}-\had_pe^{i\nu_pt}\right) \ , 
\label{deltaxhat}
\eea
where the coupling constant is defined by
\be
s^{(p)}_m=\frac{\sqrt{N}b^{(p)}_m}{\mu_p^{1/4}} \ .
\label{coupconsts}
\ee
For the center of mass mode,
\be
s^{(1)}_m=1 \quad \nu_1=\nu \ ,
\ee
and for the breathing mode
\be
s^{(2)}_m=\frac{\sqrt{N}}{\sqrt[4]{3}}
\frac{1}
{\left(\displaystyle \sum^N_{m=1}u_m^2\right)^{1/2}}\,u_m
\quad \nu_2=\sqrt{3}\nu \ .
\ee

\section {Laser-Ion interactions}\setcounter{equation}{0}
I will now consider the interaction of a laser field with the
trapped ions.  The theory must take into acount both the internal
and vibrational degrees of freedom of the ions.  I will consider 
two types of transition between internal ionic levels: 
the familiar electric-dipole allowed (E1)
transitions and dipole forbidden electric quadrupole (E2)
transitions.  Electric quadrupole transitions have
been considered in detail by Freedhoff \cite{HSF1},\cite{HSF2}.
The reason for considering forbidden transitions is that they
have very long decay lifetimes; spontaneous emission 
will destroy the coherence of a quantum computer, and
therefore is a major limitation on the capabilities of
such devices \cite{PK},\cite{HJKLP}.  Magnetic dipole
(M1) transitions, which also have long lifetimes, 
tend only to occur between sub-levels of a 
configuration, and will therefore require long wavelength
lasers in order to excite them.  As it is necessary to
resolve individual ions in the trap using the laser,
the use long wavelengths will seriously degrade
performance.  More highly forbidden transitions
are also a possibility for use in a
quantum computer.  In particular,  there is a
octupole allowed (E3) transition at 467nm with
a theoretical lifetime of $1.325\times10^8$ sec.\cite{FW},
which has recently been observed 
at the National Physical Laboratory
at Teddington, England \cite{HAK}.

The interaction picture Hamitonians 
for electric di\-pole $(E1)$ and 
electric quad\-rupole $(E2)$ 
transitions of the m-th ion, 
 located at $\br_m$ are
\bea
\hH_I^{(E1)}&=&ie\sum_{MN}\omega_{MN}\ket{N}\bra{M}
\average{N}{\hx_i}{M}A_i(\br_m,t)e^{i\omega_{MN}t} \ , \\
\hH_I^{(E2)}&=&\frac{ie}{2}\sum_{MN}\omega_{MN}\ket{N}\bra{M}
\average{N}{\hx_i\hx_j}{M}\partial_iA_j(\br_m,t) e^{i\omega_{MN}t}\ , 
\eea
where $A_j(\br,t)$ is the $j-th$ component of the vector potential
of the laser field, $\partial_i$ denotes differentiation
along the $i-th$ direction and summation over repeated indices 
$(i,j=x,y,z)$ is implied; $\hx_i$ is the  $i-th$ component of the position operator
for the valence electron of the ion;  $\{\ket{N}\}$ is the set
of all eigenstates of the unperturbed ion and the
transition frequency is $\omega_{MN}=\omega_M-\omega_N$
where the energy of the $N-th$ state is $\hbar \omega_N$ .  

For a laser beam in a standing wave configuration (see fig. 1),
propagating along a direction specified by the unit vector $\bn$,
the vector potential and its derivative are given by the formulas
\bea
A_i(\br_m,t)&=&-\beps_i\frac{E}{i \omega}
sin\left[k\hat{\zeta}_m(t)\right]e^{i\omega t}\, +\, c.c. , \\
\partial_iA_j(\br_m,t)&=&-n_i\beps_j\frac{E}{c}
cos\left[k\hat{\zeta}_m(t)\right]e^{i\omega t} \, +\, c.c.
\label {vecpot}
\eea
In (\ref{vecpot}), I have approximated the 
laser beam as a plane wave, $\beps$ being the polarization vector, 
$E$ is the amplitude
of the electric field,  
$\omega$ is the laser frequency and $k=\omega/c$ is the wavenumber.
The operator $\hat{\zeta}_m(t)$ is the distance
between the $m-th$ ion and the plane mirror used to form the
standing wave.
 
If we restrict our consideration to just two states, $\ketone$ and $\kettwo$,
and make the rotating wave equation, the  interaction Hamiltonians
may be rewritten as follows:
\bea
\hH_I^{(E1)}&=&\hbar\Omega^{(E1)}_0
sin\left[k\hat{\zeta}_m(t)\right]e^{i(t \Delta-\phi)}\ket{1}\bra{2}
\quad +\quad h.a.\ , \\
\hH_I^{(E2)}&=&i\hbar\Omega^{(E2)}_0
cos\left[k\hat{\zeta}_m(t)\right]e^{i(t \Delta-\phi)}\ket{1}\bra{2}
\quad +\quad h.a.\ ,
\eea
where the detuning is $\Delta=\omega-\omega_{21}$ and the
Rabi frequencies are given by
\bea
\Omega^{(E1)}_0&=&
\left|\frac{eE}{\hbar}\average{1}{\hx_i}{2}\beps_i \ \right| ,
\label{Rabidip} \\
\Omega^{(E2)}_0&=&
\left|\frac{eE\omega_{21}}{2 \hbar c}
\average{1}{\hx_i\hx_j}{2}\beps_i n_j \right|\ .
\label{Rabiquad}
\eea

If the standing wave of the laser is so contrived that the
equilibrium position of the $m-th$ ion is
located at a {\em node}, i.e., the electric field strength is
zero, then 
\be
\hat{\zeta}_m(t)=l\lambda+cos\theta \hq_m(t)
\ee
where $l$ is some integer,
 $\lambda$ is the wavelength, $\theta$ is the angle between
the laser beam and the trap axis and
we have assumed that 
the fluctuations of the ions transverse to the trap axis are negligible.
In this case the two Hamiltonians become:
\bea
\hH_I^{(E1)}&=&\hbar\Omega^{(E1)}_0
kcos\theta\hq_m(t) e^{i(t \Delta-\phi+ l \pi)}\ket{1}\bra{2}
\quad +\quad h.a.\ , \\
\hH_I^{(E2)}&=&\hbar\Omega^{(E2)}_0
e^{i\left[t \Delta-\phi-(l+1/2)\pi\right]}\ket{1}\bra{2}
\quad +\quad h.a.\ , 
\eea
where we have neglected terms involving 
$\hq_m(t)^2$. It is convenient to write the
displacement of the ion in terms of the creation and
annihilation operators of the phonon modes, {\em viz.}: 
\be
kcos\theta\hq_m(t)=i\frac{\eta}{\sqrt{N}}
\sum^N_{p=1}s^{(p)}_m
\left(\ha_pe^{-i\nu_pt}-\had_pe^{i\nu_pt}\right)   ,
\label{phonons}
\ee
where $\eta=\sqrt{\hbar k^2 cos^2\theta/ \ 2 M \nu}$ is called the Lamb-Dicke
parameter.

Similarly if the standing wave is arranged so that the ion is at
an {\em antinode}, i.e.
\be
\hat{\zeta}_m(t)=\frac{(2l-1)\lambda}{2}+cos\theta \hq_m(t)
\ee
then the Hamiltonians are:
\bea
\hH_I^{(E1)}&=&\hbar\Omega^{(E1)}_0
 e^{i(t \Delta-\phi+ l \pi)} \ket{1}\bra{2}
\quad +\quad h.a.\ , \\
\hH_I^{(E2)}&=&\hbar\Omega^{(E2)}_0
kcos\theta\hq_m(t) e^{i\left[t \Delta-\phi-(l+1/2)\pi\right]}
\ket{1}\bra{2}
\quad +\quad h.a.
\eea

Thus we have two basic types of Hamiltonian:
\bea
\hH_V&=&\hbar\Omega_0
 e^{i(t \Delta-\phi_v)} \ket{1}\bra{2}\quad +\quad h.a.\ , \\
\hH_U&=&\hbar\Omega_0
kcos\theta\hq_m(t)e^{i(t \Delta-\phi_u)}
\ket{1}\bra{2}
\quad +\quad h.a. ,
\label{Hu}
\eea
where $\Omega_0$ stands for either  $\Omega^{(E1)}_0$ or 
$\Omega^{(E2)}_0$.

By changing the node to an antinode, by moving the reflecting
mirror for example, we can switch from one type of Hamiltonian
to the other.  In the first case, the laser beam will only interact
with internal degrees of freedom of the ion, while in the second
case the collective motion of the ions will be affected as well.


\section{Evaluation of the Rabi frequencies}\setcounter{equation}{0}
We can relate the matrix elements appearing in the definitions of 
the Rabi frequencies to the Einstein $A$ coefficients for the transitions.
In order to do this we will rewrite the matrix elements in terms of the
Racah tensors:
\bea
\average{1}{\hx_i}{2}\beps_i&=&\sum^1_{q=-1}
\average{1}{rC^{(1)}_q}{2}c^{(q)}_i\beps_i , \\
\average{1}{\hx_i\hx_j}{2}\beps_in_j&=&\sum^2_{q=-2}
\average{1}{r^2C^{(2)}_q}{2}c^{(q)}_{ij}\beps_in_j ,
\eea
where we have used the fact that $\beps\cdot\bn=0$.
The vectors $\bc^{(q)}$ and the second rank tensors $c^{(q)}_{ij}$
may be calculated quite easily; explicit expressions are given in the
appendix. If we assume $LS$ coupling, the states $\ketone$ and
$\kettwo$ are specified by the angular momentum quantum numbers;
thus we will use the notation $\ketone=\ket{jm_j}$ and $\kettwo=\ket{j'm'_j}$
, where $j$ is the total angular momentum quantum number and $m_j$ the
magnetic quantum number of the lower state and   $j'$ is the total angular 
momentum quantum number and $m'_j$ the
magnetic quantum number of the upper state.
Using the Wigner-Eckart theorem (ref. \cite{RDC}, section 11.4),
 the matrix elements may be rewritten as
\bea
\average{1}{\hx_i}{2}\beps_i&=& \redavone
\sum^1_{q=-1}\threej{j}{1}{j'}{-m_j}{q}{m'_j}
c^{(q)}_i\beps_i , 
\label{mateldip}\\
\average{1}{\hx_i\hx_j}{2}\beps_in_j&=&\redavtwo
\sum^2_{q=-2}\threej{j}{2}{j'}{-m}{q}{m'}
c^{(q)}_{ij}\beps_in_j ,
\label{matelquad}
\eea
the terms containing six numbers in brackets being 
Wigner $3-j$ symbols (ref. \cite{RDC}, section 5.1), 
and ${\langle 1 \|r^q{\bf C}^{(q)}\| 2 \rangle}$
being the reduced matrix element.
The Einstein A coefficients for the two levels 
are given by the expressions:
\bea
\Abarone&=&\frac{4c\alpha k^3_{12}}{3}\sum^1_{q=-1}
\left|\average{1}{r C^{(1)}_q}{2}\right|^2 \\
\Abartwo&=&\frac{c\alpha k^5_{12}}{15}\sum^2_{q=-2}
\left|\average{1}{r^2 C^{(2)}_q}{2}\right|^2 .
\eea
Using the Wigner-Eckart theorem again, these expressions reduce to
the following:
\bea
\Abarone&=&\frac{4c\alpha k^3_{12}}{3}\left|\redavone\right|^2
\sum^1_{q=-1}\threej{j}{1}{j'}{-m_j}{q}{m'_j}^2  \\
\Abartwo&=&\frac{c\alpha k^5_{12}}{15}\left|\redavtwo\right|^2
\sum^2_{q=-2}\threej{j}{2}{j'}{-m_j}{q}{m'_j}^2 . 
\eea

These coefficients are the rates for spontaneous decay from the upper
level $\ketone$ to the lower level $\kettwo$.  A simpler expression for
the total rate of spontaneous decay of $\kettwo$ to all of the sublevels
of the lower state may be found by summing these rates over all values of
$m_j$:
\bea
\Aone\equiv\sum_{m=-j}^j\bar{A}^{(E1)}_{12}&=
&\frac{4c\alpha k^3_{12}}{3(2j'+1)}\left|\redavone\right|^2  
\label{EinstAdip}\\
\Atwo\equiv\sum_{m=-j}^j\bar{A}^{(E2)}_{12}&=
&\frac{c\alpha k^5_{12}}{15(2j'+1)}\left|\redavtwo\right|^2 . 
\label{EinstAquad}
\eea
These decay rates, which are the same for all of the sublevels of the
upper level, are the quantities usually quoted in data tables. Using 
(\ref{Rabidip}-\ref{Rabiquad}), (\ref{mateldip}-\ref{matelquad}) and
(\ref{EinstAdip}-\ref{EinstAquad}), we then obtain the following formula
for the Rabi frequencies:
\be
\Omega_0=
\frac{e|E|}{\hbar \sqrt{c\alpha}}\sqrt{\frac{A_{12}}{ k^3_{12}}}\sigma ,
\label{rabi2}
\ee
where
\bea
\sigma^{(E1)}&=&
\sqrt{\frac{3(2j'+1)}{4}}
\left|\sum^1_{q=-1}\threej{j}{1}{j'}{-m_j}{q}{m'_j}c^{(q)}_i\beps_i \right| , \\
\sigma^{(E2)}&=&
\sqrt{\frac{15(2j'+1)}{4}}
\left|\sum^2_{q=-2}\threej{j}{2}{j'}{-m_j}{q}{m'_j}c^{(q)}_{ij}\beps_in_j \right|\ .
\eea

The values of these quantities will be dependent on the choice of states
of ions used for the upper and lower levels, and upon the polarization and
direction of the laser beam.  As a specific example,
we will assume that the ions are in a weak 
magnetic field, which serves to define the z-direction of quantization.
Furthermore, we will assume that the lower level $\ketone$ is the
$m_j=-1/2$ sublevel of a $^2S_{1/2}$ ground state, the nucleus having
spin zero.  The upper level for
the dipole transition is a sublevel of a  $^2P_{1/2}$
state, while for the quadrupole transition it is a sublevel of a  $^2D_{3/2}$
state.  
\bea
\Omega^{(E1)}_0&=&\frac{e|E|}{\hbar}\sqrt{\frac{\Aone}{4c\alpha k^3_{12}}} , \\
\Omega^{(E2)}_0&=&\frac{e|E|}{\hbar}\sqrt{\frac{\Atwo}{2c\alpha k^3_{12}}} .
\eea

\section{Validity of Cirac and Zoller's Hamiltonian}\setcounter{equation}{0}

Equations (\ref{phonons}) and (\ref{Hu}) give the following 
expression for the Hamiltonian for the case when the laser
standing wave is so configured that it can excite the vibration
modes of the ions:
\be
\hH_U=\frac{i\eta\hbar\Omega_0}{\sqrt{N}}
\sum_{p=1}^N s^{(p)}_m\left(\ha_pe^{-i\nu_pt}-\had_pe^{i\nu_pt}\right)
e^{i(t \Delta-\phi_u)}
\ket{1}\bra{2}
\quad +\quad h.a. ,
\label{Hu2}
\ee
In their paper \cite{CZ}, Cirac and Zoller assumed   that the laser
can interact with only the center of mass mode of the ions' fluctuations.
This interaction forms a vitally important element
of their proposed method for implementing a quantum controlled
not logic gate.
Thus they used a Hamiltonian of the following form 
[c.f. ref.\cite{CZ}, eq. (1)]
\be
\hH^{(CZ)}_U=\frac{i\eta\hbar\Omega_0}{\sqrt{N}}
 \left(\ha_pe^{-i\nu_1t}-\had_pe^{i\nu_1t}\right)
e^{i(t \Delta-\phi_u)}
\ket{1}\bra{2}
\quad +\quad h.a. 
\label{Hcz}
\ee
This is an approximate form of (\ref{Hu2}), in which all of the 
other ``extraneous'' phonon modes have been neglected.  We will now
investigate under what circumstances these modes may be ignored.

We will assume that the wavefunction for a single ion interacting
with the laser beam may be written as follows:
\be
\ket{\Psi (t)}=\alpha_0(t)\ket{1}\ket{vac}+b_0(t)\ket{2}\ket{vac}+
\sum^N_{p=1}\alpha_p(t)\ket{1}\ket{1_p}+\sum^N_{p=1}b_p(t)\ket{2}\ket{1_p} ,
\ee
where $\ket{1}$ and  $\ket{2}$ are the energy eigenstates of the $m-th$ ion's
internal degrees of freedom, $\ket{1_p}$ is the state of the ions' collective vibration
in which the $p-th$ mode has been excited by one quantum, and $\ket{vac}$
is the vibrational ground state.  To avoid ambiguity, the ket for the ion's internal
state appears first, the ket for the vibrational state second.  

The equation of motion for this wavefunction is 
\be
i\hbar\frac{\partial}{\partial t}\ket{\Psi (t)}=\hH_U\ket{\Psi (t)} .
\ee
Using (\ref{Hu2}), and assuming that one cannot excite states with 
two phonons,
one obtains the following equations:
\bea
\dot{\alpha}_0&=
&\frac{\Omega_0\eta}{\sqrt{N}}\sum_{p=1}^N s_m^{(p)}\beta_p(t)\\
\dot{\beta}_0&=
&\frac{\Omega_0\eta}{\sqrt{N}}\sum_{p=1}^N s_m^{(p)}\alpha_p(t)\\
\dot{\alpha}_p&=&-i(\nu_p-\nu_1)\alpha_p
-\frac{\Omega_0\eta}{\sqrt{N}}s_m^{(p)}\beta_0(t)\\
\dot{\beta}_p&=&-i(\nu_p+\nu_1)\beta_p
-\frac{\Omega_0\eta}{\sqrt{N}}s_m^{(p)}\alpha_0(t)
\eea
We have assumed that $\Delta=-\nu_1$, so that the laser is tuned to
the specific sideband resonance required to perform
Cirac and Zoller's universal gate operation (ref.\cite{CZ}, eq. 3), namely
the two level transition $\ket{1}_n\ket{1_1}\leftrightarrow \ket{2}_n\ket{vac}$.

Since $|\alpha_0(t)|, |\beta_0(t)|\leq 1$, we can consider the following upper limits
on the amplitudes of the states which include excitation of ``extraneous'' 
phonon modes (i.e. phonon modes other than the center of mass mode):
\be
|\alpha_p(t)|\leq|A_p(t)| , \quad |\beta_p(t)|\leq |B_p(t)|
\ee
where
\bea
\dot{A}_0+i(\nu_p-\nu_1)A_p&=&-\frac{\Omega_0\eta}{\sqrt{N}}s_m^{(p)}\\
\dot{B}_0+i(\nu_p+\nu_1)B_p&=&-\frac{\Omega_0\eta}{\sqrt{N}}s_m^{(p)}.
\eea
Solving these equations one finds that
\bea
|A_p(t)|&\leq &\frac{2\Omega_0\eta}{\sqrt{N}(\nu_p-\nu_1)}|s_m^{(p)}|\\
|B_p(t)|&\leq &\frac{2\Omega_0\eta}{\sqrt{N}(\nu_p+\nu_1)}|s_m^{(p)}| ,
\eea
Thus the total probability that ``extraneous'' modes are excited
has the following upper limit:
\be
P_{ext}=\sum_{p=2}^N|\alpha_p(t)|^2+|\beta_p(t)|^2
\leq2\left(\frac{2\Omega_0\eta}{\sqrt{N}\nu}\right)^2
\left[\sum_{p=2}^N\frac{\mu_p+1}{(\mu_p-1)^2}(s_m^{(p)})^2 \right],
\ee
where we have used the definition of the mode frequencies (\ref{nup})
and the fact that the eigenvalue for the center of mass mode is $\mu_1=1$.
This quantity will be different for each ion in the string; taking its average
value, we find
\bea
\bar{P}_{ext}&\equiv&\frac{1}{N}\sum_{m=1}^NP_{ext} \nn \\
&\leq&2\left(\frac{2\Omega_0\eta}{\sqrt{N}\nu}\right)^2 \Sigma(N) ,
\eea
where we have used the definition of the coupling constants,
(\ref{coupconsts}) and the orthonormality of the eigenvectors
(\ref{beeorth}).  The function $ \Sigma(N)$ is defined by
the formula
\be
\Sigma(N)=\sum_{p=2}^N\frac{\mu_p+1}{(\mu_p-1)^2\sqrt{\mu_p}} .
\label{siggy}
\ee
This must be evaluated numerically by solving for the eigenvalues of
the trap normal modes for different numbers of trapped ions $N$. 
The results are shown in fig. (3).  The function varies slowly with
the value of $N$, and , for $N\geq10$, we can, to a good approximation,
replace it by a constant $\Sigma(N)\approx 0.82$. 
Thus we obtain the following upper limit
on the total probability of the ``extraneous'' phonon modes
becoming excited:
\be
\bar{P}_{ext}\leq\left(\frac{2.6\,\Omega_0\eta}{\sqrt{N}\nu}\right)^2 .
\ee
Thus we obtain the following sufficiency condition for the validity
of Cirac and Zoller's Hamiltonian (\ref{Hcz}):
\be
\left(\frac{2.6\,\Omega_0\eta}{\sqrt{N}\nu}\right)^2 \ll 1 .
\ee
\section {Conclusion}\setcounter{equation}{0}
In this preceeding sections we have reviewed the
theoretical basis for cold trapped ion quantum
computation.  How these various laser-ion interaction
effects may be
combined to perform fundamental quantum logic gates
is descibed in the seminal work of Cirac and Zoller \cite{CZ}.
Using the formulae given here one can determine,
for example, the laser field strength required or
the separation between ions in the trap.  Such things are
of great importance in the engineering of
practical devices.

Finally there is the question of what type of ion to use.
Figure 4 show the energy levels of four suitable species
of ion.  These have been chosen based two criteria: that
the lowest excited state has a forbidden transition to the 
ground state, and their popularity amoungst published
ion trapping experiments.  It is not intended that this is
an exhaustive list of suitable ions, but rather it is
to show the properties of typical ions.  

\section*{Acknowledgments}
The author would like to thank Barry Sanders (Macquarie
University, Australia) and
Ignacio Cirac (University of Innsbruch, Austria) 
for useful discussions and Albert Petschek (Los Alamos
National Laboratory, USA) for reading an earlier version of the
manuscript.
This work was funded by the 
National Security Agency.

\section*{Appendix}
\renewcommand{\theequation}{{\rm
A.\arabic{equation}}}\setcounter{equation}{0}

The vectors $c^{(q)}_i$  are usual  normalized spherical basis vectors:
\bea
\bc^{(1)}&=&-\frac{1}{\sqrt{2}}(1,-i,0) , \\
\bc^{(0)}&=&(0,0,1) , \\
\bc^{(-1)}&=&\frac{1}{\sqrt{2}}(1,i,0) . \\
\eea
Note that
\bea
\bc^{(q)}&=&(-1)^q\bc^{(-q)\ast} \\
\bc^{(q)}\cdot \left\{\bc^{(q')}\right\}^\ast&=&\delta_{q,q'} .
\eea

The second rank tenors $c^{(q)}_{ij}$ are given by the formula
\be
c^{(q)}_{ij}=\sqrt{\frac{10}{3}}(-1)^q\sum_{m_1,m_2=-1}^1
\threej{1}{1}{2}{m_1}{m_2}{-q}c^{(m_1)}_{i} c^{(m_2)}_{j}  .
\ee
Explicity these five tensors are:
\bea
c^{(2)}_{ij}&=&\frac{1}{\sqrt{6}}
\left(\begin{array}{ccc}
1&-i&0\\
-i&-1&0\\
0&0&0
\end{array}\right) , \\
c^{(1)}_{ij}&=&\frac{1}{\sqrt{6}}
\left(\begin{array}{ccc}
0&0&-1\\
0&0&i\\
-1&i&0
\end{array}\right) , \\
c^{(0)}_{ij}&=&\frac{1}{3}
\left(\begin{array}{ccc}
-1&0&0\\
0&-1&0\\
0&0&2
\end{array}\right) , \\
c^{(-1)}_{ij}&=&\frac{1}{\sqrt{6}}
\left(\begin{array}{ccc}
0&0&1\\
0&0&i\\
1&i&0
\end{array}\right) , \\
c^{(-2)}_{ij}&=&\frac{1}{\sqrt{6}}
\left(\begin{array}{ccc}
1&i&0\\
i&-1&0\\
0&0&0
\end{array}\right) .
\eea
Note that
\bea
c^{(q)}_{ij}&=&(-1)^qc^{(-q)\ast}_{ij} \\
\sum_{ij}c^{(q)}_{ij}\left\{c^{(q')}_{ij}\right\}^\ast&=
&\frac{2}{3}\delta_{q,q'} .
\eea

\newpage
\section*{Figure Captions}
\noindent
Figure 1.  A schematic diagram of ions in 
a linear trap, to illustrate the notation used in this paper.

\vspace{10mm}
\noindent
Figure 2.  The relationship between 
the number of trapped ions $N$
and the minimum separation.  
The curve is given by (\ref{separations}) 
while the points come from the numerical solutions of
the algebraic equations (\ref{ionequs}).

\vspace{10mm}
\noindent
Figure 3.  The function $\Sigma(N)$ defined by (\ref{siggy}).

\vspace{10mm}
\noindent
Figure 4.  Energy level diagrams for four species of ions suitable
for quantum computation.  Wavelengths and lifetimes are given for
the important transitions, the numbers in square brackets being
the reference for the data.  The lifetime is the reciprocal of the
Einstein A coefficient defined in (\ref{EinstAdip}) and 
(\ref{EinstAquad}).  The thick lines are dipole
allowed (E1) transitions, the thin lines quadrupole allowed (E2)
transitions.  The atomic number and the mass number
of the most abundant isotope (with its relative abundance) are also given.
Since all of these isotopes have an even mass number, they do
not have a nuclear spin.

\vspace{20mm}
\section*{Table Captions}
\noindent
Table 1. Scaled equilibrium positions of the trapped ions, for different total
numbers of ions.  This data was obtained by numerical solutions of
(\ref{ionequs}).  The length scale is given by (\ref{eleanor}).

\vspace{10mm}
\noindent
Table 2.  Numerically determined eigenvalues and 
eigenvectors of the matrix $A_{nm}$
defined by (\ref{alice}), for 2 to 10 ions.
The eigenvectors are
normalized as defined by (\ref{beeorth}).


\begin{thebibliography}{99}
\bibitem{GPA}Accurate values for the total lifetimes 
of the P states of $Ca^+$ are given in 
R. N. Gosselin, E. H. Pinnington and W. Ansbacher:
{\em Phys. Rev. A} {\bf 38}, 4887 (1988); 
the lifetimes of the S-P transitions can be 
calculated using this data and the data from the 
NBS tables \cite{WSM}.
\end{thebibliography}
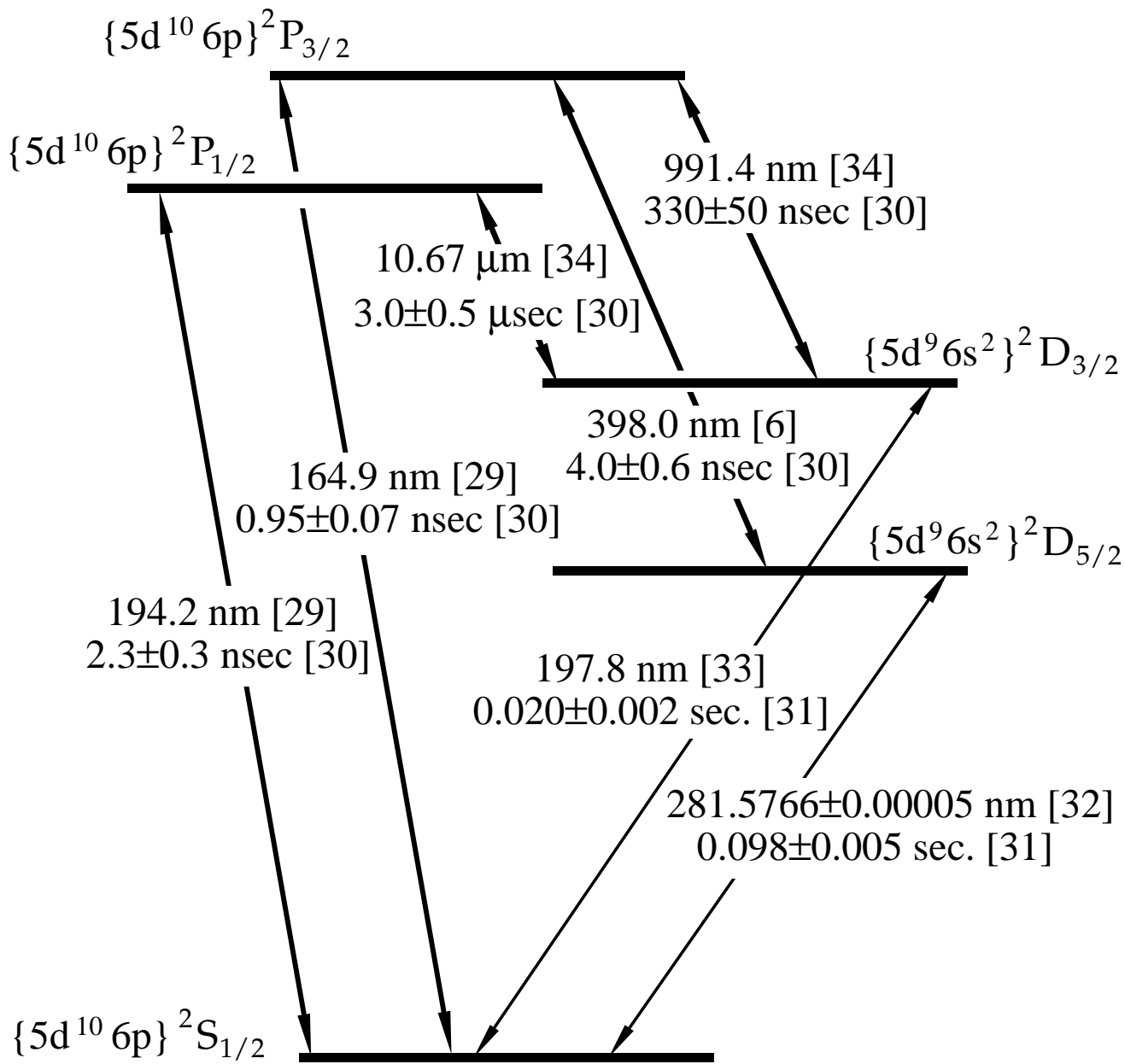
\end{document}